\pdfoutput=1
\documentclass[universe,article,submit,pdftex,moreauthors]{mdpi} 
\usepackage{caption}
\usepackage{subcaption}

\preto{\abstractkeywords}{\nolinenumbers}
\firstpage{1} 
\pubvolume{1}
\issuenum{1}
\articlenumber{0}
\pubyear{2023}
\copyrightyear{2023}
\datereceived{ } 
\daterevised{ } 
\dateaccepted{ } 
\datepublished{ } 
\hreflink{https://doi.org/} 

\newcommand{\aj}{Astrophysical Journal}
\newcommand{\apj}{Astrophysical Journal}
\newcommand{\apjl}{Astrophysical Journal Letters}
\newcommand{\apjs}{Astrophysical Journal Supplement}
\newcommand{\araa}{Annual Review of Astronomy and Astrophysics}
\newcommand{\jcap}{Journal of Cosmology and Astroparticle Physics}
\newcommand{\mnras}{Monthly Notices of the Royal Astronomical Society}
\newcommand{\nat}{Nature}
\newcommand{\pasa}{Proceedings of the Astronomical Society of Australia}
\newcommand{\pasj}{Proceedings of the Astronomical Society of Japan}
\newcommand{\pasp}{Proceedings of the Astronomical Society of the Pacific}
\newcommand{\physrep}{Physics Reports}
\newcommand{\prd}{Physical Review D}
\newcommand{\prl}{Physical Review Letters}

\Title{Kilonova-Targeting Lightcurve Classification for Wide Field Survey Telescope}

\TitleCitation{Kilonova-Targeting Lightcurve Classification for Wide Field Survey Telescope}

\Author{Runduo Liang$^{1,2,}$*\orcidA{}, Zhengyan Liu$^{1,2}$\orcidB{}, Lei Lei$^{2,3}$\orcidC{} and Wen Zhao $^{1,2}$\orcidD{}}

\AuthorNames{Runduo Liang, Zhengyan Liu, Lei Lei and Wen Zhao}

\AuthorCitation{Liang, R, D.; Liu, Z, Y.; Lei, L.; Zhao W.}

\address{%
$^{1}$ \quad CAS Key Laboratory for Researches in Galaxies and Cosmology, Department of Astronomy, University of Science and Technology of China, Chinese Academy of Sciences, Hefei, Anhui 230026, China\\
$^{2}$ \quad School of Astronomy and Space Sciences, University of Science and Technology of China, Hefei 230026, China\\
$^{3}$ \quad Purple Mountain Observatory, Chinese Academy of Sciences, Nanjing 210023, China}

\corres{Correspondence: aujust@mail.ustc.edu.cn}

\abstract{With the enhancement of sensitivity of  Gravitational Wave (GW) detectors and capabilities of large survey facilities, such as Vera Rubin Observatory Legacy Survey of Space and Time (LSST) and 2.5-m Wide Field Survey Telescope (WFST), we now have the potential to detect an increasing number of distant kilonova (KN). However, distinguishing KN from the plethora of detected transients in ongoing and future follow-up surveys presents a significant challenge. In this study, our objective is to establish an efficient classification mechanism tailored for the follow-up survey conducted by WFST, with a specific focus on identifying KN associated with GW. We employ a novel temporal convolutional neural network architecture, trained using simulated multi-band photometry lasting for 3 days by WFST, accompanied by contextual information, i.e. luminosity distance information by GW. By comparison of the choices of contextual information, we can reach 95\% precision, and 94\% recall for our best model. It also performs good validation on photometry data on AT2017gfo and AT2019npv. Furthermore, we investigate the ability of the model to distinguish KN in a GW follow-up survey. We conclude that there is over 80\% probability that we can capture true KN in selected 20 candidates among $\sim 250$ detected astrophysical transients that have passed real-bogus filter and cross-matching. }

\keyword{Gravitational wave astronomy; Neutron stars; Gravitational wave sources; Transient detection} 

\begin{document}

\section{Introduction} \label{sec:intro}

The merger of binary neutron stars (BNS) and neutron star-black hole (NSBH) could be the source of thermal emission extended 
from near-infrared to ultraviolet, which is powered by r-process generated radioactive decay of heavy elements in the ejecta during the merger, often referred as kilonova (KN)~\citep{2016PhRvL.116f1102A,2017ApJ...848L..12A,2017PASA...34...69A}. It is believed that 
kilonova are typically fainter than supernova and fast fading within a week~\citep{1998ApJ...507L..59L,metzger2010electromagnetic}. 
Often during this process, a highly relativistic jet along the polar axis could launch short $\gamma-$ray bursts (sGRB) lasting for a few seconds~\citep{paczynski1991cosmological,1992ApJ...395L..83N,2007PhR...442..166N}. 
Interaction of the jet and interstellar medium powers X-ray afterglow which is spread within a relatively wide viewing angle~\citep{2002ApJ...576..120T,2011ApJ...736L..21R,2012ApJ...746...48M,2014ApJ...784L..28N}. The dawn of multimessenger astronomy associated with gravitational wave (GW), heralded by the detection of compact binary merger, GW170817, and its 
electromagnetic (EM) counterparts AT2017gfo, GRB170817A, 
has ushered in a new era of multimessenger astrophysics~\citep{2017ApJ...848L..12A,2017PASA...34...69A,2017ApJ...848L..19C,2017Sci...358.1570D,2017Sci...358.1583K,2017PASJ...69..101U,2017ApJ...848L..32M,2017ApJ...848L..18N,2017Sci...358.1574S,2017ApJ...848L..16S,2017Natur.551...67P,2017Natur.551...75S,2017ApJ...848L..27T,2018ApJ...855L..23A}. 
With the completion of the thereafter third LIGO/Virgo observing run, O3, and a total of 90 GW candidates identified, 
GW astrophysics has jumped into a time-domain era~\citep{2021PhRvX..11b1053A,2021arXiv211103606T}. 
In addition, the observation of EM signature has allowed for more accurate inclination and distance measurements of the host galaxy by model fitting and identification. In fact, EM counterparts of GWs are important sources of bright standard sirens to 
constrain the Hubble constant, e.g. $H_{0}=74^{+16}_{-8} \rm km\ s^{-1}\ Mpc^{-1}$ for GW170817, which could shed light on the Hubble tension problem~\citep{2017Natur.551...85A,Dietrich_2020}.

In the span of O3 run, a total of 56 public alerts were released by LIGO/Virgo through Gamma-ray Coordinate Network (GCN) 
Notices and Circulars\footnote{\url{https://emfollow.docs.ligo.org/userguide/}}~\citep{2021PhRvX..11b1053A}. While extensive prompt follow-up observations were conducted following low latency public alerts which yield hundreds to 
thousands of candidates, no confirmed kilonvoa was identified~\citep{kasliwal_kilonova_2020,andreoni_growth_2020}. 
The reason for the undesirable outcome is controversial. The fast fading signature of KN and very limited sky coverage induced by poor localization might be 
responsible~\citep{saguescarracedo_detectability_2021,zhu_no_2021} However, poor sky coverage and selection criteria are 
also effective to data stream, which indicates that  true KN could be rejected by real-bogus classification, 
astrophysical origin selection and even KN classification~\citep{andreoni_growth_2020}. 

The undergoing LIGO/Virgo/KAGRA (LVK) fourth observing run, O4, will reach $\sim 160$ Mpc for BNS merger detection 
and over $\mathcal{O}(10)$ are expected to detect by LVK~\citep{2018LRR....21....3A,2023arXiv230609234W}. 
And the sensitivity of O5 run will extend to $\sim$330 Mpc detection in the next decade, 
which implies the potential for discovering BNS merger will increase by an order of magnitude.~\citep{2022ApJ...924...54P}.
Many efforts are focusing on rapid deployment and optimization of optical follow-up triggered 
by GW and GRB public alerts~\citep{2022ApJ...927..163C}, e.g. Zwicky Transient Facility\footnote{\url{https://www.ztf.caltech.edu/}}  (ZTF,~\citep{2022ApJS..260...18A,kasliwal_kilonova_2020}), 
Dark Energy Camera\footnote{\url{https://www.darkenergysurvey.org/}} (DECam,~\citep{2022ApJ...927...50R,andreoni_growth_2020}), Wide Field Survey Telescope\footnote{\url{https://wfst.ustc.edu.cn/}} (WFST,~\citep{2023arXiv230607590W,2023ApJ...947...59L}) 
and next generation Vera Rubin Observatory Legacy Survey of Space and Time\footnote{\url{https://www.lsst.org/}} (LSST,~\citep{2017ApJ...846...62S,coughlin_optimizing_2018}). 

The 2.5-m WFST, installed at Saishiteng Mountain near Lenghu on Tibetan Plateau, China, will strongly support various science cases including time-domain astronomy, asteroids and the solar system, the Milky Way and its satellite dwarf galaxies, galaxy formation and cosmology and so on~\citep{2023ApJ...947...59L}.
With a field of view (FoV) of 6.55 $\rm deg^{2}$, it could cover $\sim 10^{3}\ \rm deg^{2}$ within a night with 5$\sigma$ depth of 
22.31, 23.42, 22.95, 22.43, 21.50 mag under 30s exposure in five bands ($u,g,r,i,z$) respectively, making it one of the most powerful
facilities in the northern sky for discovering GW counterparts~\citep{2023RAA....23c5013L}. In addition, excellent observing conditions with an average seeing of $0.75$ arcsec and 22.0 $\rm mag\ \rm arcsec^{-2}$ V-band background provide potential for high quality data~\citep{2021Natur.596..353D}.

It is common to be overwhelmed by the data stream produced by the rapid and deep searching of wide field instruments. Since it is not 
sufficient to identify KN by photometry solely, efficient KN classification is still of great significance for maximizing identified KN. 
Traditionally, the KN photometry classification is based on several criteria, e.g. decay rate, color 
evolution~\citep{kasliwal_kilonova_2020} or model fitting and it is upgraded to a complete pipeline, e.g. ZTFReST~\citep{2021ApJ...918...63A}. 
Another way is employing a machine learning classifier, which was implemented during O3 and is well designed so far~\citep{andreoni_growth_2020}.
Stachie et al.~\citep{2020MNRAS.497.1320S} adapted the long-term lightcurve RAPID~\citep{2019PASP..131k8002M} classifier  short-term KN detections. 
Chatterjee et al.~\citep{chatterjee_-cid_2021} deployed a KN classifier, which uses similar structure, including GW skymap information. 
Biswas et al.~\citep{biswas_enabling_2022} designed a fast transient classification algorithm, aiming to KN, which is implemented as a module in \textsc{Fink} broker\footnote{\url{https://fink-broker.org/}}, 
a data stream processor software for ZTF and LSST. Sravan et al.~\citep{2023arXiv230709213S} proposed a fully machine-directed pipeline for KN discovery including the optimizing survey plan to maximize the chance to identify KN.

Of great essence is the need for a rapid and efficient KN classifier for the WFST Target-of-Opportunity (ToO) observation. Thus in this work, we simulate lightcurve data from KN and contaminants detected by WFST with designed capability, air conditions and strategies. Subsequently, a modified RAPID framework is employed to train and test the performance on our simulated WFST ToO data generated using mock GW skymaps~\citep{2022ApJ...924...54P}.

This paper is organized as follows. Section~\ref{sec:simulation} describes the simulation of training data. We begin with the mock GW skymaps and survey plans that are generated automatically through the ToO pipeline. KN and contaminants are simulated using Monte Carlo simulations which follow their space and time distribution, i.e. KN is distributed following GW and contaminants are distributed following their volumetric rate. Then we 
implement mock photometry and collect their lightcurves and associated information like location and line-of-sight probability. The classifier framework and training process are detailed in Section~\ref{sec:classification}. In Section~\ref{sec:results}, we test the performance on both dataset, real data and explore the situation 
in simulated GW follow-up surveys. Finally, we present our conclusions in Section~\ref{sec:conclusion}.

\section{Transients Simulation}\label{sec:simulation}

\begin{figure}
    \centering
    \includegraphics[width=0.65\linewidth]{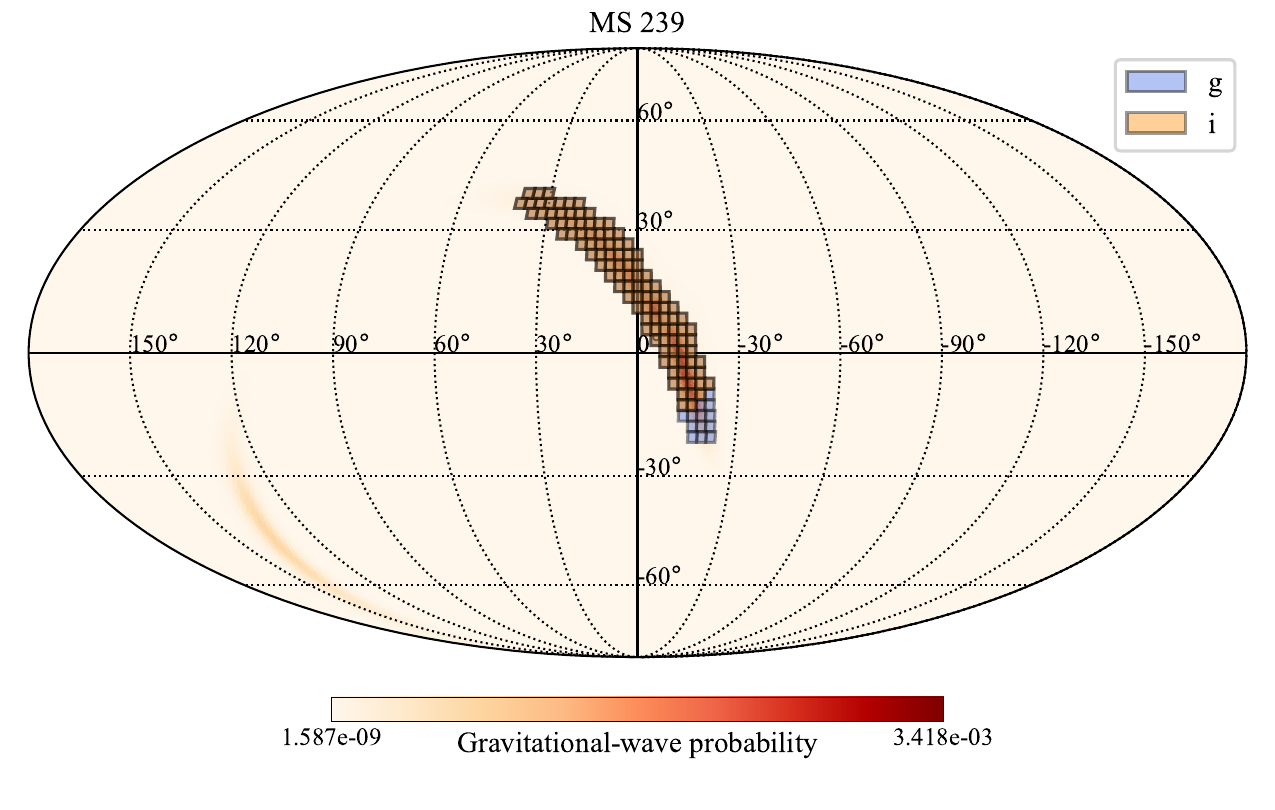}

    \includegraphics[width=0.65\linewidth]{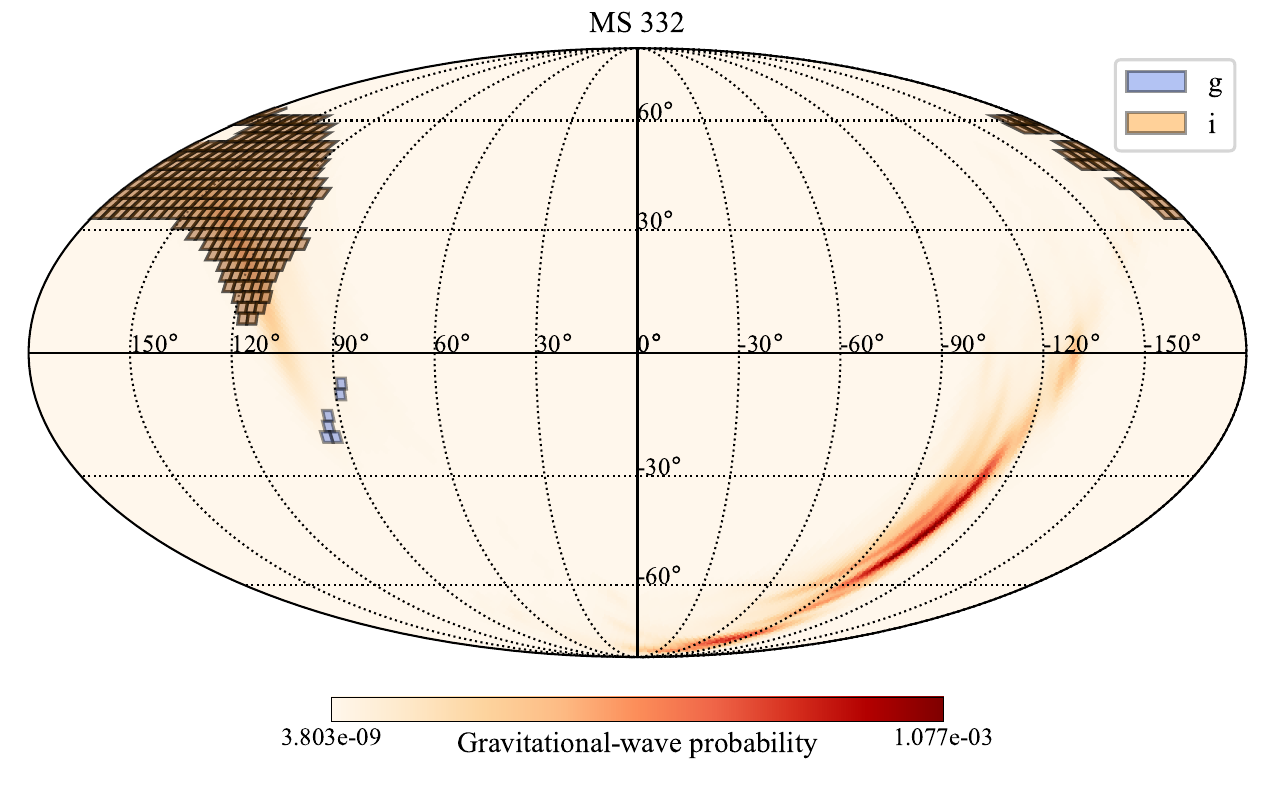}

    \caption{Examples of mock GW skymap and footprints of WFST ToO observation.}
    \label{fig:tiles}
\end{figure}

Our simulation has three steps: firstly we randomly choose mock GW events detected by LVK 
with O4 sensitivity. For each event, many KN and contaminant objects are placed across the sky. Then, we trigger our KN ToO pipeline to deploy mock surveys. Finally, we collect lightcurves given certain survey plans and contextual information for each object. The full dataset is obtained by iterating 
all select events.

Since the aim of this work is the KN classifier for WFST during O4, the amount of BNS and NSBH events detected so 
far is not adequate to cover the diversity of skymaps and survey cadences. 
Petrov et al.~\citep{2022ApJ...924...54P} simulated GW signals from BBH, BNS and NSBH merger and 
employed a more realistic threshold under O3, O4, O5 sensitivity. They gave comparable sky localization to O3 and concluded that the sky localization might be even worse during O4, which coincides with recent observations. Therefore, we randomly choose 250 GW events\footnote{\url{https://zenodo.org/records/4765750}} 
from BNS merger as our mock events that will happen at random in 2024, 
in which the WFST will be online. 

\subsection{Survey Simulation}

In practical GW observations, an accurate localization of the event plays a pivotal role in triggering ToO observations. To initiate it effectively, it is essential that the observable area for the event aligns with the survey capability of the telescope, ensuring a sufficiently high probability of observing the specific sky region. In our study, we carefully exclude sections of the localization area falling within $\pm 15^{\rm o}$ of the galactic plane and regions with declination $\delta<-30^{\rm o}$ for the mock GW skymaps. Additionally, we incorporate the restriction that the airmass should not exceed 2 to define the observable sky area for each event. Based on the observable area and the probability of observing the sky region, we filter 250 events, selecting those that meet the following criteria: an observable area is less than 3000 $\rm deg^{2}$ and the probability of observing the 2D sky region is not less than 0.2. Among them, 68 events finally trigger ToO survey. 

Then, we generate survey plans by implementing \texttt{gwemopt}\footnote{\url{https://github.com/skyportal/gwemopt}}, which was originally developed by \citet{coughlin_optimizing_2018}, serving the purpose of optimizing the EM follow-up search for GW events. During O3, several post-event observation plans were formulated using \texttt{gwemopt} for both individual and joint observations by multiple telescopes~\citep[e.g.,][]{coughlin_growth_2019,kasliwal_kilonova_2020,antier_grandma_2020,anand_optical_2021,frostig_infrared_2022}, yielding good performance. 

In the process of generating survey plans, \texttt{gwemopt} includes several algorithms within each step: (1) Skymap tiling, (2) Time allocations, (3) Scheduling. \citet{coughlin_optimizing_2018} extensively discussed the efficiency of various combinations of these algorithms, determining that the combination of the (1) MOC algorithm, (2) power-law algorithm, and (3) greedy algorithm yielded the most efficient results. Therefore, we employ this combination of algorithms for our simulations. Specific inputs need to be prepared before running the code, namely merger event, exposure time, bands, and observation windows. Initially, the merger times for the GW events were randomly distributed from 2024.01.01 to 2024.12.31 to match the future operational timeline of O4. The selection of nightly coverage frequencies and bands was based on the observable area of the events and the lunar phase conditions. $g,i-$ and $r,i-$bands were prioritized around the new moon and full moon, respectively. We also adjust our exposure time according to the luminosity distance of the event and estimated time spent in observation. Currently, we assume an observation window for three days post-merger, during which WFST would repeatedly cover the target area. Upon the completion of the code, it will generate a list containing the corresponding pointings for each exposure, the observation time, and the cumulative probability within the coverage area as output. We show two examples of GW skymap, MS\_239 and MS\_332, and corresponding triggered WFST ToO survey in Figure~\ref{fig:tiles}. The tiles in the map are footprints of one-night survey. 

\begin{figure*}[htbp]
    \centering
    \includegraphics[width=0.8\linewidth]{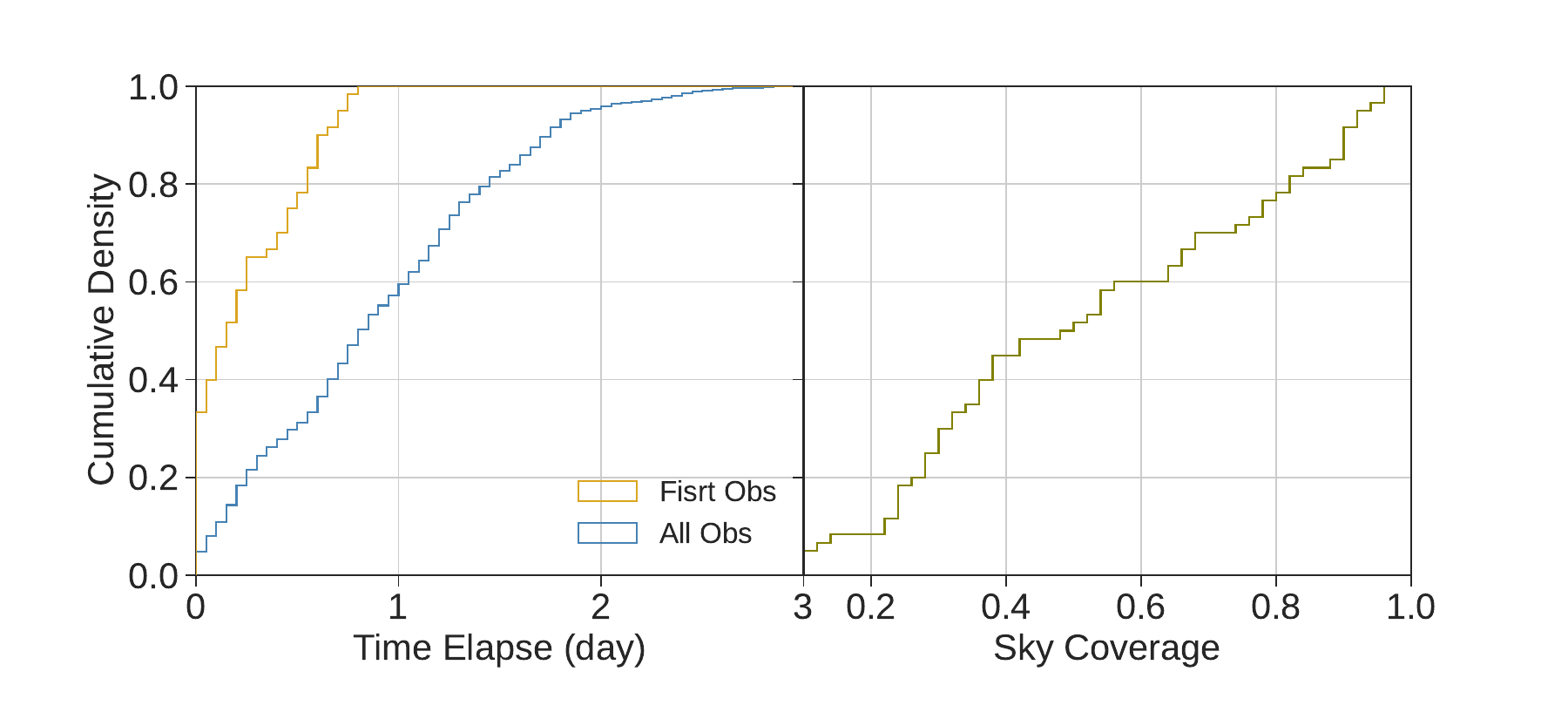}
    \caption{Left panel: Cumulative density of time elapse of follow-up first and all observation. Right panel: Cumulative density of sky coverage of triggered events by WFST.}
    \label{fig:coverage}
\end{figure*}

In Figure~\ref{fig:coverage}, we show the cumulative density of elapsed time between follow-up trigger and GW trigger. The first observation time and overall observations are almost uniform within 0.8 days and 2 days respectively. The sky coverage is also widely distributed between $0.2-1$, indicating that we have covered as many scenarios as we can. 

\subsection{KN Simulation}\label{subsec:kn}

For KN simulation, we use two models to generate Spectral Energy Distribution (SED). The first one is a two-component model first presented 
by Bulla~\citep{2019MNRAS.489.5037B,Dietrich_2020}, in which the spectra of KN were calculated using Monte Carlo radioactive 
transfer code \texttt{POSSIS}\footnote{\url{https://github.com/mbulla/kilonova_models}}. The first component, dynamical ejecta, is characterized by mass $M_{\rm dyn}$, which have velocities ranging from $0.08c \sim 0.3c$. There are two compositions that are formed via different channels, in which lanthanide-rich 
composition is distributed within angle $\pm \Phi$ around the equatorial plane which is formed due to tidal processes and lanthanide-free composition otherwise that is formed by hydrodynamic interaction. The second component is post-merger wind ejecta, $M_{\rm pm}$, 
which is distributed spherically and has relatively lower velocities ranging from $0.025c \sim 0.08c$ representing outflow from the accretion disk.
To generate SEDs of arbitrary parameters, we construct a surrogate model using a Neural Network following the method in Ref.~\citep{2017PhRvD..96l3011D,2021arXiv211215470A,2022arXiv220508513P}. The ejecta mass and velocity are related to the binary property involving mergers, such as chirp mass, mass 
ratio and equation of state (EoS). Many numerical relativity simulations have analyzed the process of merger and KN explosions
~\citep{sekiguchi_dynamical_2016,dietrich_modeling_2017,radice_binary_2018,foucart_remnant_2018,coughlin_multimessenger_2019,nedora_mapping_2022}. Therefore, 
it allows us to bridge a connection between BNS sample to a KN lightcurve sample. 

For dynamical ejecta $M_{\rm dyn}$, we use fitting formula from~\citet{coughlin_multimessenger_2019} which was extended from Ref.~\citep{radice_binary_2018,2018MNRAS.480.3871C},

\begin{equation}
\small
\label{bns_m_dyn}
\log _{10} M_{\mathrm{dyn}}^{\mathrm{fit}}= {\left[a \frac{\left(1-2 C_1\right) M_1}{C_1}+b M_2\left(\frac{M_1}{M_2}\right)^n+\frac{d}{2}\right] }+(1 \leftrightarrow 2),
\end{equation}
where $M_{1,2}$ and $C_{1,2}$ represent the mass and compactness of two compact objects respectively and fitting constants $a=-0.0719,\ b=0.02116,\ c=-2.42,\ d=-2.905$. $(1 \leftrightarrow 2)$ represents the exchanging of subscript. 
The mass of dynamical ejecta is sensitive to mass ratio $q=M_{1}/M_{2}$ and compactness of neutron stars. We also characterize the fraction of the red component of ejecta presented in Ref.~\citep{2016PhRvD..93l4046S,Nicholl:2021rcr} that separate lanthanide-rich and -poor components by a threshold 
$Y_{e}\sim 0.25$ when $q>0.8$, whereas for $q<0.8,\ f_{\mathrm{red}}\sim 1$ because the less massive NS is disrupted by tidal forces that suppress the shock~\citep{2013ApJ...773...78B,2013PhRvD..87b4001H,2016CQGra..33r4002L,2017CQGra..34j5014D}. 
The fraction of the red component can be 
written as

\begin{equation}
	\label{fred}
	f_{\mathrm{red}} = \mathrm{min} \left( 1,\ a q^2 + b q + c \right),
\end{equation}
with $a=14.8609,\ b=-28.6148,\ c=13.9597$. Using the spherical ejecta density profile assumption in the Bulla model, it is easy to obtain the half-opening angle for lanthanide-rich component. 

For post-merger ejecta, we use the following expression to evaluate disk mass~\citep{coughlin_multimessenger_2019},

\begin{equation}
	\centering
	\footnotesize
	\log _{10}\left(M_{\mathrm{disk}}\right)=\max \left(-3, a\left(1+b \tanh \left(\frac{c-M_{\mathrm{tot}} / M_{\mathrm{thr}}}{d}\right)\right)\right),
\end{equation}
with fitting parameters that include mass ratio dependence $a=a_{0}+\delta a\cdot \xi,b=b_{0}+\delta b\cdot \xi$ and the free parameter $\xi$ given by 
\begin{equation}
	\small
	\xi=\frac{1}{2} \tanh \left(\beta\left(q-q_{\text {trans }}\right)\right),
\end{equation}
where $q=M_{1}/M_{2}\leq1$. The best-fit values of free parameters are $a_{0}=-1.581,\delta a=-2.439,b_{0}=-0.538,\delta b=-0.406,c-0.953,d=0.0417,\beta=3.910,\hat{q}_{\mathrm{trans}}=0.900$. 
$M_{\mathrm{tot}}$ is the total mass of BNS and $M_{\mathrm{thr}}$ represents the threshold mass for prompt massive neutron star collapse to black hole with the expression~\citep{Dietrich_2020}

\begin{equation}
	\centering
	\small
	M_{\mathrm{thr}}=\left(2.38-3.606 \frac{M_{\mathrm{TOV}}}{R_{1.6}}\right) M_{\mathrm{TOV}},
\end{equation}
where $M_{\mathrm{TOV}}$ is the maximum stable mass for non-rotating NS and $R_{1.6}$ represents the radii of $1.6 M_{\odot}$ NS.
The disk wind ejecta mass is assumed to be proportional to disk mass $M_{\mathrm{pm}}=f\cdot M_{\mathrm{disk}}$ with $f$ ranging from $0.1\sim0.5$~\citep{2015MNRAS.446..750F,2017PhRvL.119w1102S,2019MNRAS.482.3373F,2019MNRAS.490.4811C}. In this work, we fix $f=0.2$ as default set. Noted that explicit density, heating rate and opacity profile, and structured jet are considered in the new version of \texttt{POSSIS 2.0} model~\citep{2023MNRAS.520.2558B,2023arXiv230314277S}. 

We also employ another semi-analytical kilonova model presented in~\citet{Nicholl:2021rcr}, thereafter MOSFiT KN, which is implemented by the open-source software \texttt{MOSFiT}\footnote{\url{https://github.com/guillochon/mosfit}}. The model is based on the binary property forward to SED. They employed numerical relativity fitting formula to convert binary parameters to ejecta parameters. The ejecta contains 
three components with different grey opacity, namely blue ($\kappa=0.5\ \rm cm^{2}g^{-1}$), red ($\kappa=10\ \rm cm^{2}g^{-1}$), purple ($\kappa=0.5\ \rm cm^{2}g^{-1}$) components. The red and blue components represent dynamical 
ejecta with higher velocity and the half opening angle of red ejecta is fixed to $\Phi=45^{\rm o}$. This model also includes the cocoon emission and magnetic enhancement on blue ejecta mass. 

Once the KN model is prepared, we can obtain the KN SED sample from a BNS sample and inject their parameters 
into two KN models. In the standard isolated binary formation channel, a recycled NS is born first and spins up due to the accretion or recycling process. It is accompanied by a non-recycled NS that spins down after birth. 
The mass of recycled NS follows a two-Gaussian distribution,

\begin{equation}
	\centering
	\small
	\pi\left(m \mid \mu_1, \sigma_1, \mu_2, \sigma_2, \alpha\right)=\frac{\alpha}{\sigma_1 \sqrt{2 \pi}} \times \exp \left[-\left(\frac{m-\mu_1}{\sqrt{2} \sigma_1}\right)^2\right]+\frac{1-\alpha}{\sigma_2 \sqrt{2 \pi}} \exp \left[-\left(\frac{m-\mu_2}{\sqrt{2} \sigma_2}\right)^2\right]
\end{equation}
with $\mu_1=1.34\ M_{\odot},\mu_2=1.47\ M_{\odot},\sigma_1=0.02\ M_{\odot},\mu_2=0.05\ M_{\odot}\ \rm and \ \alpha=0.68\ M_{\odot}$. For non-recycled NS, they are found to follow a uniform distribution within $1.15\sim 1.42\ M_{\odot}$~\citep{2019ApJ...876...18F}. 
Given the mass of the neutron star, we calculate the radius and compactness by sampling the parameterized EoS obtained by~\citet{Dietrich_2020}, where the EoS sample was calibrated with constraints of pulsars and multimessenger observation on GW170817. We generate the KN SED sample with the size of $\sim 10^{4}$ for each KN model.

\begin{figure}[htbp]
    \centering
    \includegraphics[width=0.5\linewidth]{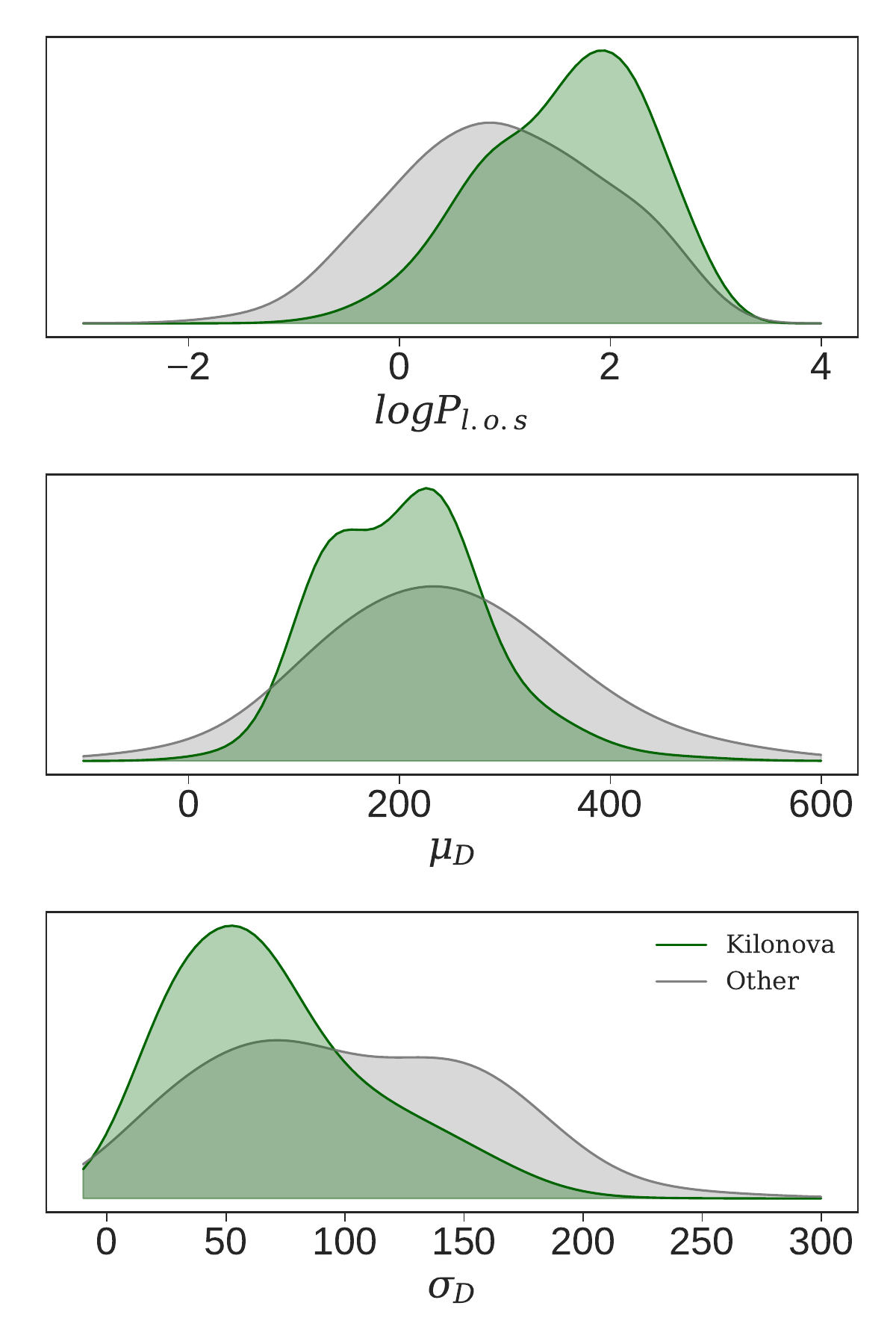}
    \caption{Distribution of line-of-sight probability, distance mean and standard deviation of the training set.}
    \label{fig:statistic}
\end{figure}

\subsection{Other Transients Simulation}

To simulate contaminants, our main focus in this work is on supernovae, one of the most significant types of contaminants following catalog cross-matching. Still, we cannot ignore other important fast transients, e.g. Cataclysmic Variables (CVs), afterglows and shock breakouts. They will be included in the complete pipeline with the aid of other selection procedures. 

\textit{SN Ia.} The Type Ia Supernova is thought to be powered by the re-ignition and thermalnuclear explosion of carbon-oxygen white dwarfs once they exceed their critical mass~\citep{1986ARA&A..24..205W,2004ApJ...612L..37P}. It is the most common and numerous contaminant object 
to KN. To model 
SN Ia, we use the SED library presented in~\citet{2007ApJ...663.1187H}. Apart from classic SN Ia, some subgroups in the SN Ia were identified through decays of observations. SN1991bg-like (SN Ia-91bg) stands out as one of the most important potential contaminants due to their bright luminous, rest-frame $m_{\rm B}\gtrsim-18$ and fast evolution feature, lightcurve width less than 70\% average SN Ia~\citep{1992AJ....104.1543F,1993ApJ...413L.105P}. 

\textit{SN Ibc.} The stripped envelope supernova explosion, also referred as to Type Ib and Type Ic supernova, 
characterizes the feature of lacking helium in spectra~\citep{2016MNRAS.461L.117E}. The lightcurves of SN Ibc are fainter, redder and evolve slower, indicating they are sub-dominant sources of contaminant. We use the SED template presented in~\citet{2002PASP..114..803N}\footnote{\url{https://c3.lbl.gov/nugent/nugent_templates.html}} to model SN Ibc, SN Ia and its subgroups.

\textit{SN II.} Type II Supernova are explosions of massive stars with mass $8\lesssim M \lesssim 18\ M_{\odot}$
~\citep{2009MNRAS.395.1409S}. They are distinguished from other types by the presence of hydrogen in their spectra. We model 
two subgroups SN IIn and SN IIP using the SED template in~\citet{2002PASP..114..803N}. 

\textit{SLSN-I.} Type I Superluminous Supernova is one of the brightest optical transients with peak absolute 
magnitudes $\lesssim-21\ \rm mag$. They are widely distributed in metal-poor dwarf host galaxies, of which some are 
powered by magnetar with very strong magnetic field~\citep{2014ApJ...787..138L}. Their spectra tend to be blue and lack hydrogen features and brighten rapidly, 
thus making them the main contaminant in early KN identification~\citep{2011ApJ...743..114C,2011Natur.474..487Q}. 
To model SED, we employ the extended library of 960 SEDs of \texttt{MOSFiT slsn} model by~\citet{2019PASP..131i4501K}.

\subsection{Training Set}

Given GW skymaps and corresponding survey strategies and transient models described above, we perform survey simulation implemented by \texttt{simsurvey}\footnote{\url{https://github.com/ZwickyTransientFacility/simsurvey}}~\citep{2019JCAP...10..005F}, a software for survey simulation and lightcurve collection~\citep{Feindt_2019,kasliwal_kilonova_2020}. The simulated KN for each GW event is sampled randomly based on two KN models in Sec~\ref{subsec:kn}. The explosion time of KN is set the same as the GW event and its location is produced following the probability density of GW skymap. 
For contaminants, we assume a uniform distribution of RA and Dec within the observed field of survey and specific redshift model listed in Tabel~\ref{table:dataset}. 

\begin{table}[htbp]
\newcolumntype{C}{>{\centering\arraybackslash}X}
\caption{Summary of transients in the training set.}
\label{table:dataset}
\begin{tabularx}{\textwidth}{CCC}
\toprule
Object Type & Number Count & Rate $(\rm yr^{-1}\rm Mpc^{-3})$\\ \midrule
Bulla KN    & 19694        & - \\
MOSFiT KN   & 20459        & - \\
SNIa        & 13553        & $3\times 10^{-5}(1+z)$ \\
SNIbc       & 2988         & $2.25\times 10^{-5}(1+z)$ \\
SNIIn       & 5492         & $7.5\times 10^{-6}(1+z)$ \\
SNIIP       & 926          & $1.2\times 10^{-4}(1+z)$ \\
SLSN        & 4712         & $2\times 10^{-8}$ \\
SNIa-91bg   & 9180         & $3\times 10^{-6}(1+z)^{1.5}$ \\
\bottomrule
\end{tabularx}
\end{table}

We randomly loop 60 skymaps and collect their lightcurves and object information, e.g. line-of-sight probability $P_{\rm l.o.s}$, distance mean and standard deviation $\{\mu_{D},\sigma_{D}\}$ derived from GW skymap. Noted that for supernovas we truncate the distance with $z<1$ which is comparable to detection depth of WFST. Finally, we obtain the data set containing 
77,005 objects and details are listed in Table~\ref{table:dataset}. Since we aim to distinguish true KN among many 
contaminants, we label Bulla KN and MOSFiT KN as \textit{Kilonova} and the rest as \textit{Other}. Figure~\ref{fig:statistic} shows the normalized distribution of line-of-sight probability, mean and standard deviation of luminosity distance of transient according to GW skymap of our dataset per class. One can see that the KN has more tight distribution of features compared with contaminants because they always happen in the high probability area of GW skymap. Besides the lightcurve, these features could also assist us in classifying KN and contaminants. 

\section{Binary Classification} \label{sec:classification}

We employ temporal convolutional network (TCN) architecture~\citep{2018arXiv180301271B} which is implemented in \texttt{keras}\footnote{\url{https://github.com/philipperemy/keras-tcn}} and \texttt{RAPID}\footnote{\url{https://github.com/daniel-muthukrishna/astrorapid}} for long-term lightcurve classification revealing good performance. We use a similar architecture to ~\citet{chatterjee_-cid_2021} who took modification to \texttt{RAPID} for binary classification to filter KN.

\begin{figure}[htbp]
    \centering
    \includegraphics[width=0.73\linewidth]{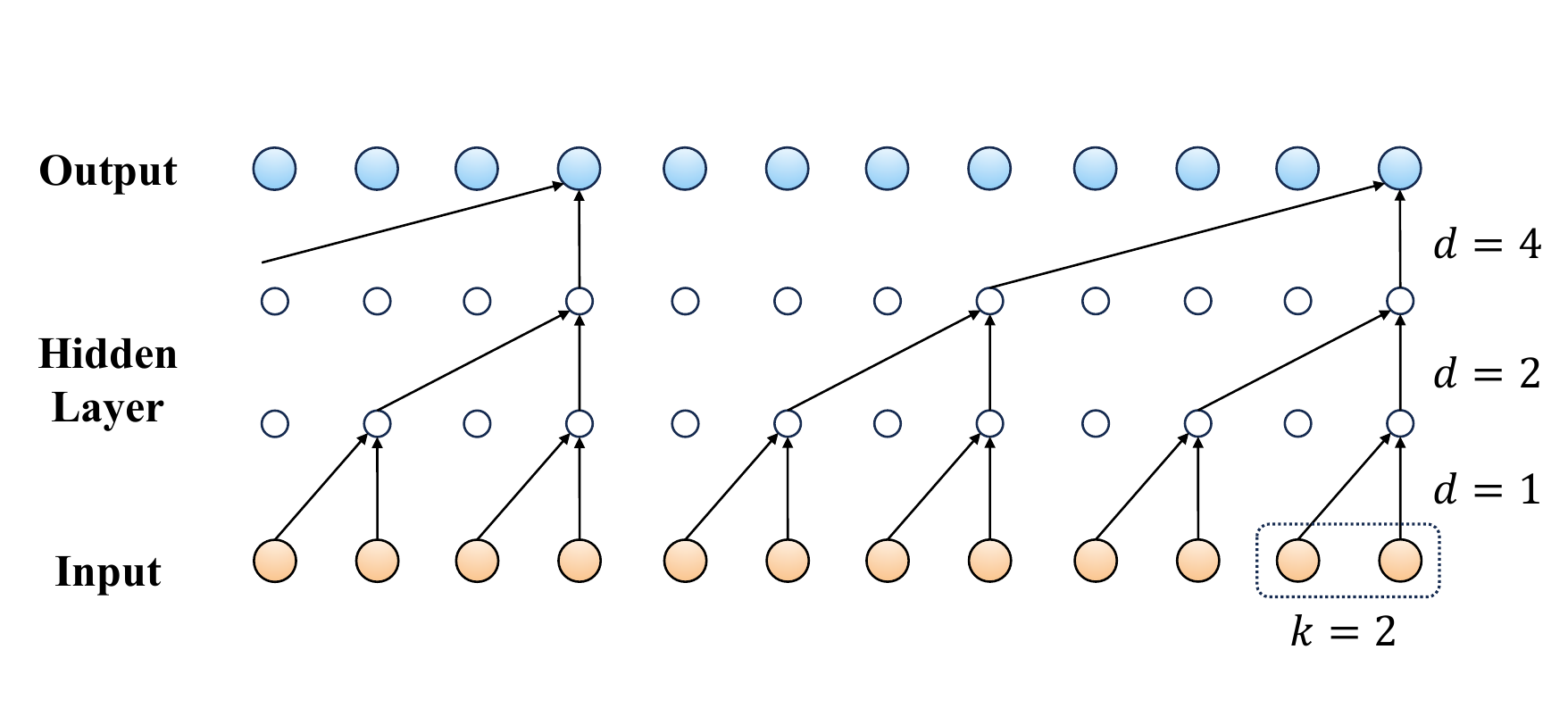}
    \caption{Framework of TCN with length of \textit{filter} is 2 and dilation is $(1,2,4)$ in hidden layers.}
    \label{fig:tcn}
\end{figure}

\subsection{TCN Framework}

The strength and utility of the TCN architecture in the classification of time series lie in its ability to promptly update results as new data becomes available. As introduced by~\citet{2018arXiv180301271B} in 2018, the TCN architecture operates on two fundamental principles: (1) The length of the output sequence matches that of the input, and (2) the internal convolutions are causal in nature. Provided that a sequence $\{ x_{1}, x_{2}, \cdots, x_{N} \}$ of data, the TCN outputs a sequence $\{ y_{1}, y_{2}, \cdots, y_{N} \}$, and the convolutions within the hidden layers of the architecture are designed in such a way that each output, $y_{n}$, solely relies on the information within the input sequence $\{ x_{1}, x_{2}, \cdots, x_{n} \}$, where n ranges from 1 to N. Within this architectural framework, as shown in Figure~\ref{fig:tcn}, one can control the extent to which long-term historical information influences the output by making judicious selections of kernel sizes or incorporating \textit{dilated} layers. Beyond that, one can inject contextual information as input, which stays unchanged with time. In this work, we consider line-of-sight probability, mean and standard deviation of luminosity distance of transient according to GW skymap and the combination of them. For simulated WFST surveys with time interval between two photometries $\sim 1$ day, in order to reveal more precise shape of lightcurve, we adopt linear interpolation with time interval $\sim 0.5$ day within 5 days. Therefore, the dimensions of input data matrix are $N\times(p+n)$, where $N,\ p,\ n$ represents the length of interpolated time series, number of passbands and number of contextual information. 

\subsection{Training}

\begin{figure}
    \centering
    \includegraphics[width=0.45\linewidth]{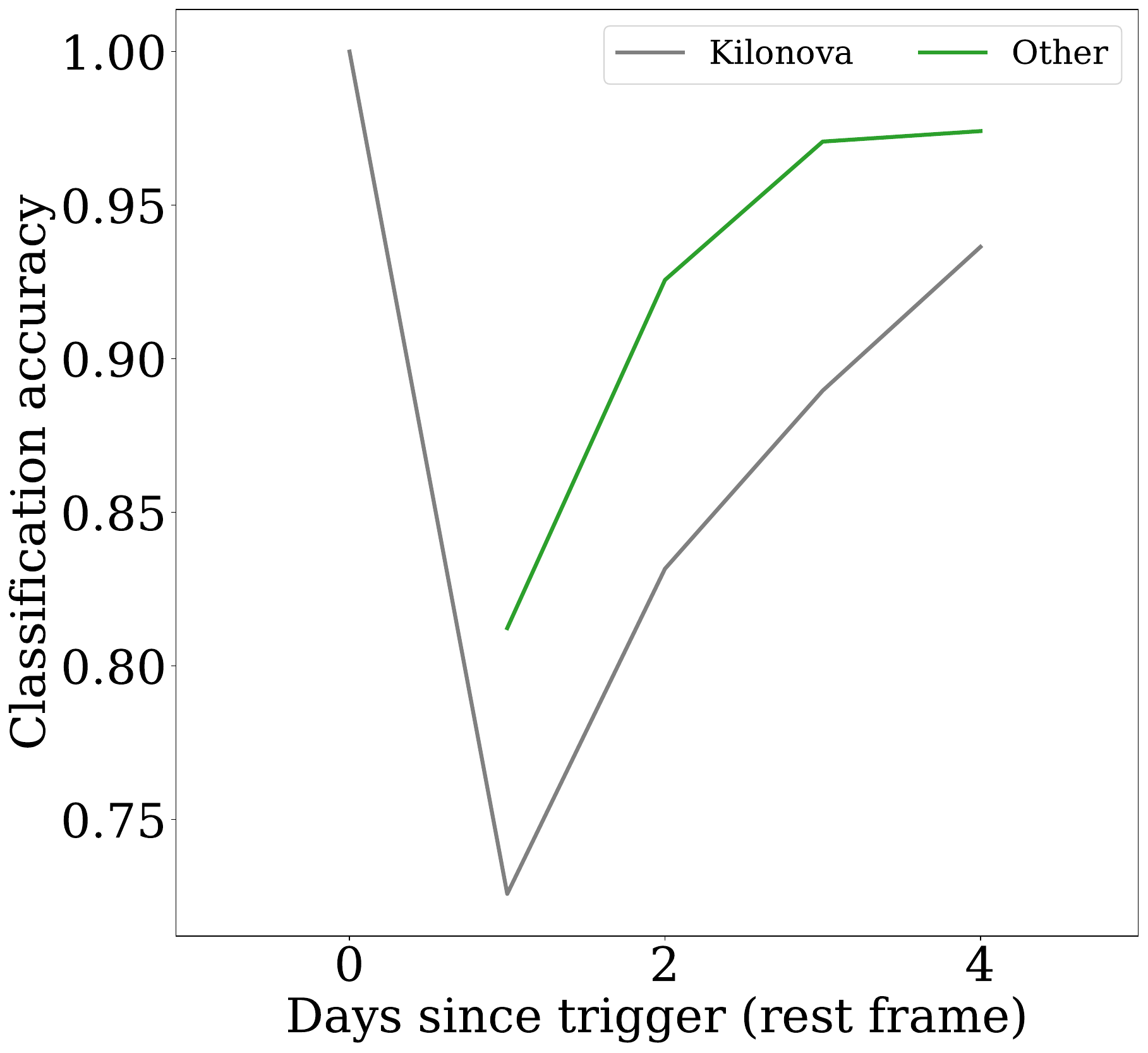}
    \caption{Accuracy per class with time of PMD model. \textit{Kilonova} and \textit{Other} reach over 98\% and 92\% accuracy after 3 days since trigger respectively, after the inaccurate prediction with high KN score when lacking photometry data.}
    \label{fig:acc}
\end{figure}

For our purpose of classification for KN in ToO data, which lasts within 3 days post-merger as shown in Figure~\ref{fig:coverage}, we use a network of filter length $k=2$ and dilation layers $d=(1,2)$ and 2 stacks to deepen the network. We employ categorical cross-entropy as the loss function in conjunction with the Adam optimizer. For dataset, we have partitioned it into a training set comprising 70\% of the data and a test set containing the remaining 30\%. By testing, we find that loss tends to converge after 50 epochs of training so that we train models for 50 epochs. Approximately 2-3 hours will be allocated for training and testing models. Notably, we find that the computational cost for training is comparatively modest, negating the necessity for hardware optimizations such as GPU acceleration. Figure~\ref{fig:acc} shows the accuracy per class with time in which at least 3 days of observation yields good accuracy. The classifier tends to predict a high KN score inaccurately in instances where photometry data is lacking, as evidenced by a noticeable decline around one day. However, as more photometry data is processed, the predictions converge more closely toward the actual label.

\begin{figure*}[htbp]
    \centering
    \includegraphics[width=0.99\linewidth]{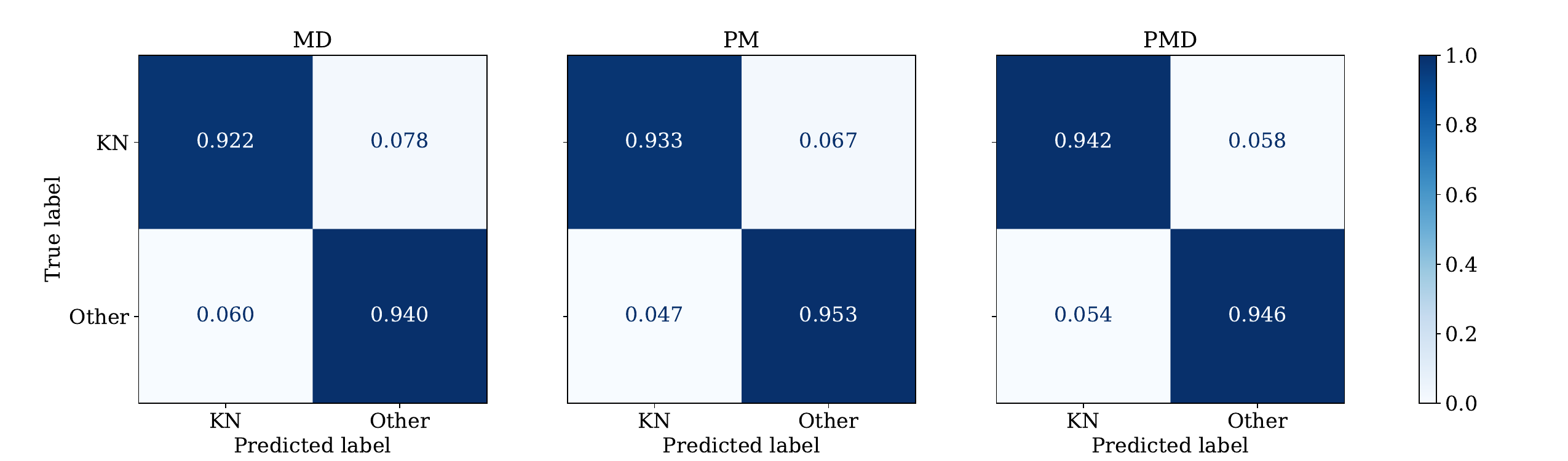}

    \includegraphics[width=1\linewidth]{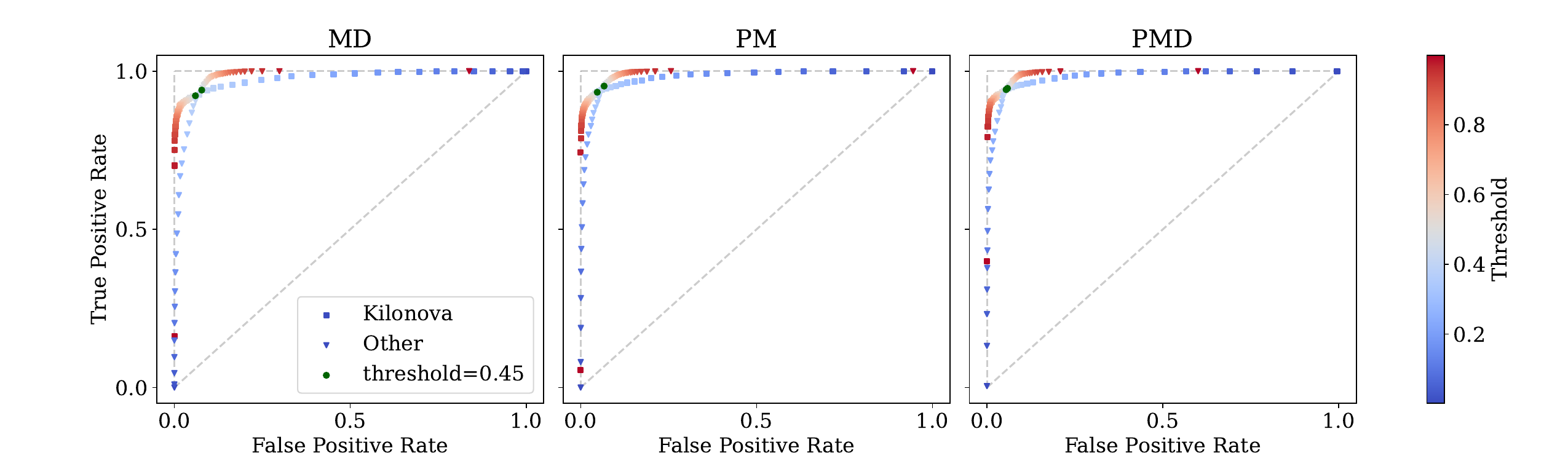}
    \caption{Upper panel: Confusion metrics for MD, PM and PMD model. The value in elements are normalized by the size of true labels per class. Lower panel: ROC curves with \textit{KN} and \textit{Other} are drawn as inverted triangles and squared lines. The colormap of marker denotes the various values of threshold and the default value of 0.45 is plotted as green dots. }
    \label{fig:cf}
\end{figure*}

As described above, we have three contextual information at our disposal to assist in classification, line-of-sight Probability, thereafter $P$, Mean and standard Deviation of luminosity distance along the line of sight of transients according to GW, hereafter $M$ and $D$. To see which information affects the result most and to decide the best combination of information used, we consider three combinations of them, MD, PM and PMD. Our fundamental principle relies on the mean luminosity distance defined by the GW skymap, serving as crucial guidance for the AB magnitude of transients. Therefore, MD model and PM model lack line-of-sight probability and measurement error of luminosity distance whereas PMD has full contextual information. 

\section{Results}\label{sec:results}

\subsection{Performance on Dataset}

Using the dataset simulated in Section~\ref{sec:simulation} and the TCN framework described in Section~\ref{sec:classification} we obtain MD, PM and PMD classification models for WFST targeting KN. Among them, PMD model reaches an overall accuracy of 98.41\%. To test the performance of a classifier, it is natural to employ a confusion matrix to show their capability. In our binary classification issue, the dimension of confusion matrix is $2\times 2$ with true positives and true negatives residing in diagonal elements while false positives and false negatives lies in upper right and lower left element respectively. Figure~\ref{fig:cf} illustrates the confusion matrices for the MD, PM, and PMD models. The matrix values are normalized by the number of true labels and the KN score is the last KN probability in the sequence of prediction. In the context of prioritizing the detection of KN, a threshold of 0.45 is selected to minimize the potential for false positives. Specifically, we classify a candidate as a KN when the predicted KN probability in the final epoch surpasses this specified threshold. Upon comparing these models, it becomes evident that the MD model exhibits lower efficiency when contrasted with the other two alternatives. 

However, it is obviously unreasonable to use the same threshold for three different models. We estimate the performance of models under various thresholds. In the low panel of Figure~\ref{fig:cf}, we show the receiver operating characteristic curve (ROC) which illustrates the diagnostic ability of a binary classifier system. The inverted triangle curve and squared curve represent true positive rate and false positive rate with threshold ranging from $10^{-3}\sim 1-10^{-3}$. The threshold of 0.45 is shown as green dots in each curve. The more powerful the model is, the closer the curve is to the upper left of the coordinate. We summarize their precision and recall with thresholds corresponding to the cross of two curves, yielding 
94.151\% precision, 94.028\% recall for PM model and 94.830\% precision, 94.125\% recall for PMD model. We notice that PM model and PMD model have comparable capability while lacking line-of-sight probability is the most intolerable case. 

\begin{figure}
    \centering
    \begin{subfigure}[b]{0.49\textwidth}
    \centering
    \includegraphics[width=0.95\linewidth]{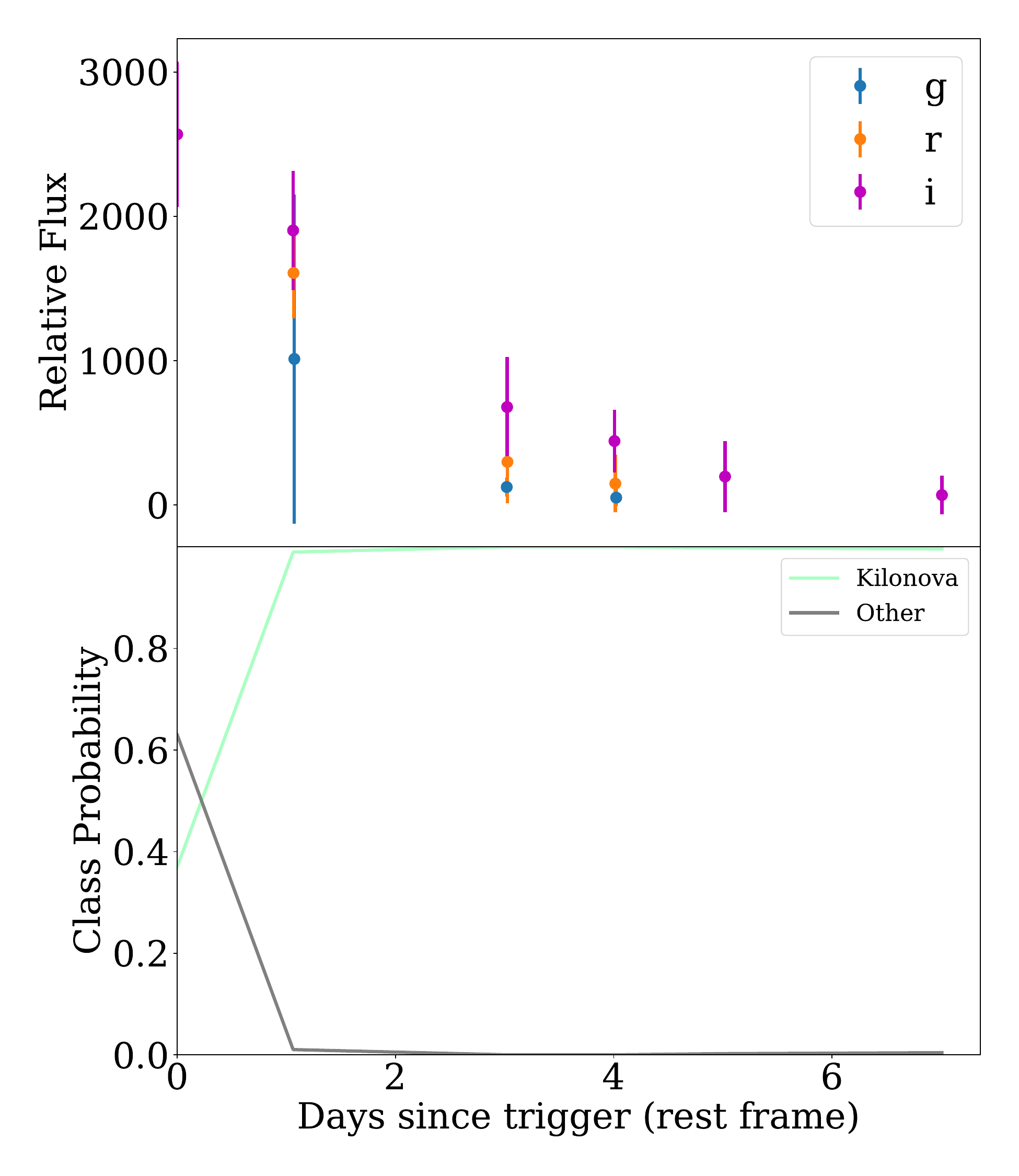}
    \end{subfigure}
    \hfill
    \begin{subfigure}[b]{0.49\textwidth}
    \centering
    \includegraphics[width=0.95\linewidth]{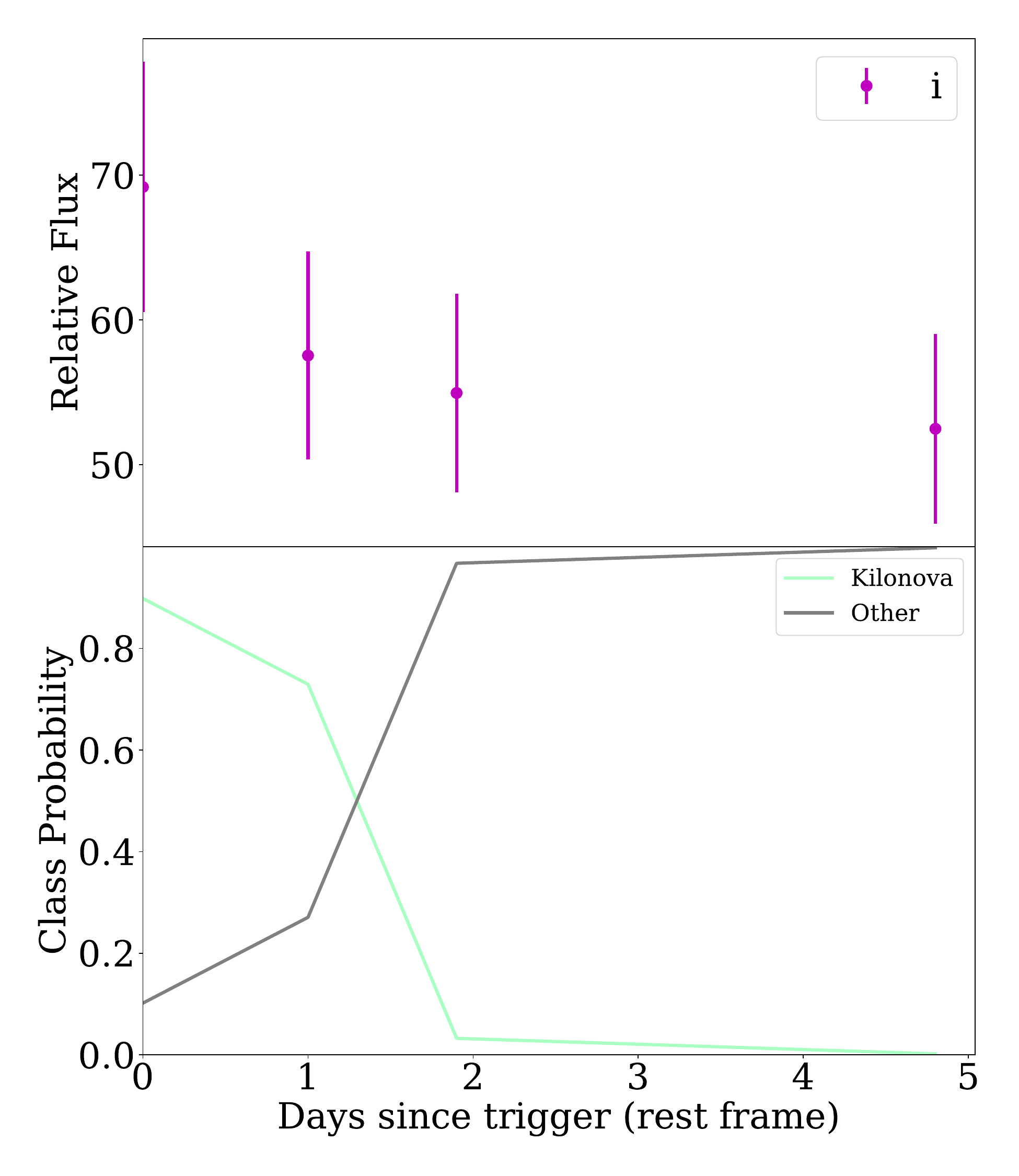}
    \end{subfigure}
    \caption{Left panel: Prediction of Swope observations on At2017gfo. Right panel: Prediction of DECam observations on AT2019npv. }
    \label{fig:real_data}
\end{figure}

\subsection{Performance on Real Data}

To further investigate the performance of the classifier, we employ similar procedures in~\citet{chatterjee_-cid_2021} to test in unseen data. We consider Swope observations on AT2017gfo~\citep{2017Sci...358.1556C}, originally identified as 'SSS17a' by the Swope team at the time of its discovery. It was the only identified KN accompanied by GW as far, with ejecta mass $\sim 0.05 M_{\odot}$ by the fitting of various KN models~\citep{2017ApJ...848L..18N,2018ApJ...855L..23A,Dietrich_2020}. We use $g-,r-,i-$band data to be compatible with our trained model input. Although the optical transmission curves of Swope are different from WFST, we maintain the belief that these disparities in the KN score's uncertainty will not lead to misleading classification results. Additionally, we also consider a counterpart candidate of GW190814~\citep{2020ApJ...896L..44A}, AT2019npv, which was a SN Ibc but identified as KN in the early phase by efforts of several teams because it was located in GW skymap and the redshift was consistent with the distance information of GW~\citep{2020MNRAS.492.3904A,2020ApJ...901...83M,andreoni_growth_2020}. We include first week $i-$band observations from DECam as WFST survey would not exceed 5 days.  

In Figure~\ref{fig:real_data}, we implement our PMD model to predict KN score of AT2017gfo and AT2019npv. We acquire comparable results to~\citet{chatterjee_-cid_2021}, in which confidence KN score is obtained from AT2017gfo and persuasive primary KN score for AT2019npv at early phase than decisive contaminants result with the increasing observations. 

\subsection{Performance on Mock Survey}

\begin{figure}
    \centering
    \includegraphics[width=0.88\linewidth]{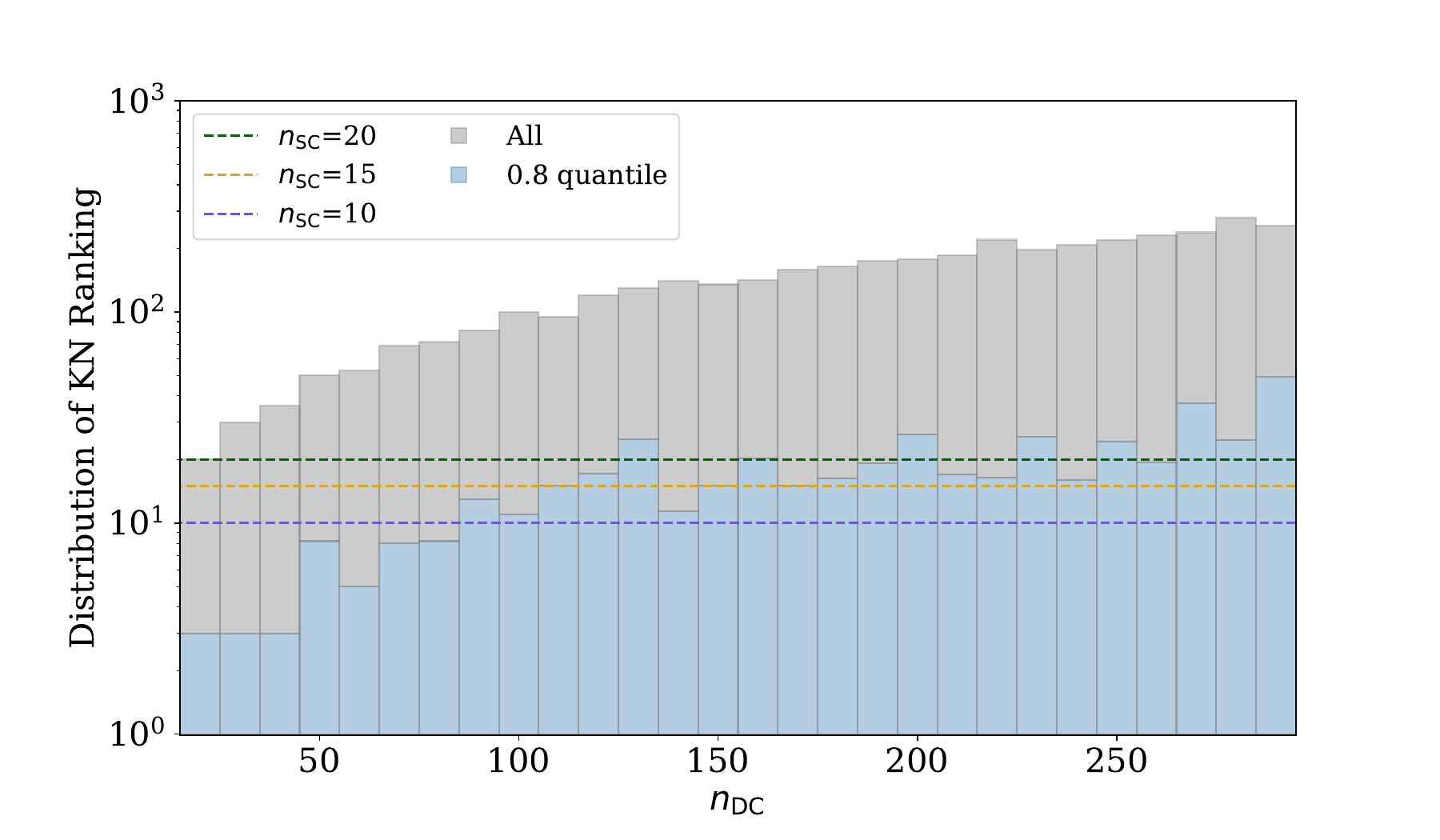}
    \caption{Distribution of ranking of true KN in data package with various $n_{\rm DC}$. The top of grey and blue bars represent maximum and 0.8 quantile of KN rankings with various $n_{\rm DC}$. }
    \label{fig:ranking}
\end{figure}

We have conducted thorough validation and performance tests of our model as described above. Nevertheless, it remains imperative to assess the model's efficiency in predicting future survey data, ensuring its practical applicability. To achieve that, we apply simulations with mock surveys described in Section~\ref{sec:simulation}. In a night survey, the WFST would detect $\mathcal{O}(100)-\mathcal{O}(1000)$ transients ignoring history activity~\citep{2023ApJ...947...59L,2023arXiv230810545L}. We assume that approximately a fraction of them can be filtered out by cross-matching or angular offset to their host galaxy. Therefore, we only consider cases of processing less than 300 candidates by classifier. As a comparison, an average of 170 EM candidates were filtered after multi-step machine learning, which includes real-bogus test, cross matching and history exclusion~\citep{kasliwal_kilonova_2020}. 
To simulate candidates processed by classifiers, which have been filtered by real-bogus test, cross matching and history exclusion, we simulate one KN according to GW skymap and $n$ contaminants with uniform distribution across the sky for each mock survey. Then we compile them into a data package, labeled as detected candidates. The size of detected candidates is denoted as $n_{\rm DC}$. We generate 5 data packages for each skymap for data augmentation, considering the variety of KN models and location. In overall, we generate 300 data packages and we rank KN score in descending order in each data package, i.e. $\rm rank=1$ means the true KN has the highest KN score in detected candidates. According to our pipeline, we will take several selected candidates, the size of them is denoted as $n_{\rm SC}$, for the subsequent spectra follow-up. In our analysis, we choose $n_{\rm SC}=\{ 10,15,20 \}$ and $n_{\rm DC} \in \left( 20,300 \right)$. In Figure~\ref{fig:ranking}, we show the distribution of ranking of KN with $n_{\rm DC}=\{ 20,30,40,\cdots,290 \}$. The top of grey and blue bars represent maximum and 0.8 quantile of KN rankings with various $n_{\rm DC}$, which means over $80\%$ KN would rank within 20 when $n_{\rm DC}$ do not exceed 290. And we also calculate 0.6 quantile which stays 1 with any $n_{\rm DC}$. 

\begin{figure}[htbp]
    \centering
    \includegraphics[width=0.68\linewidth]{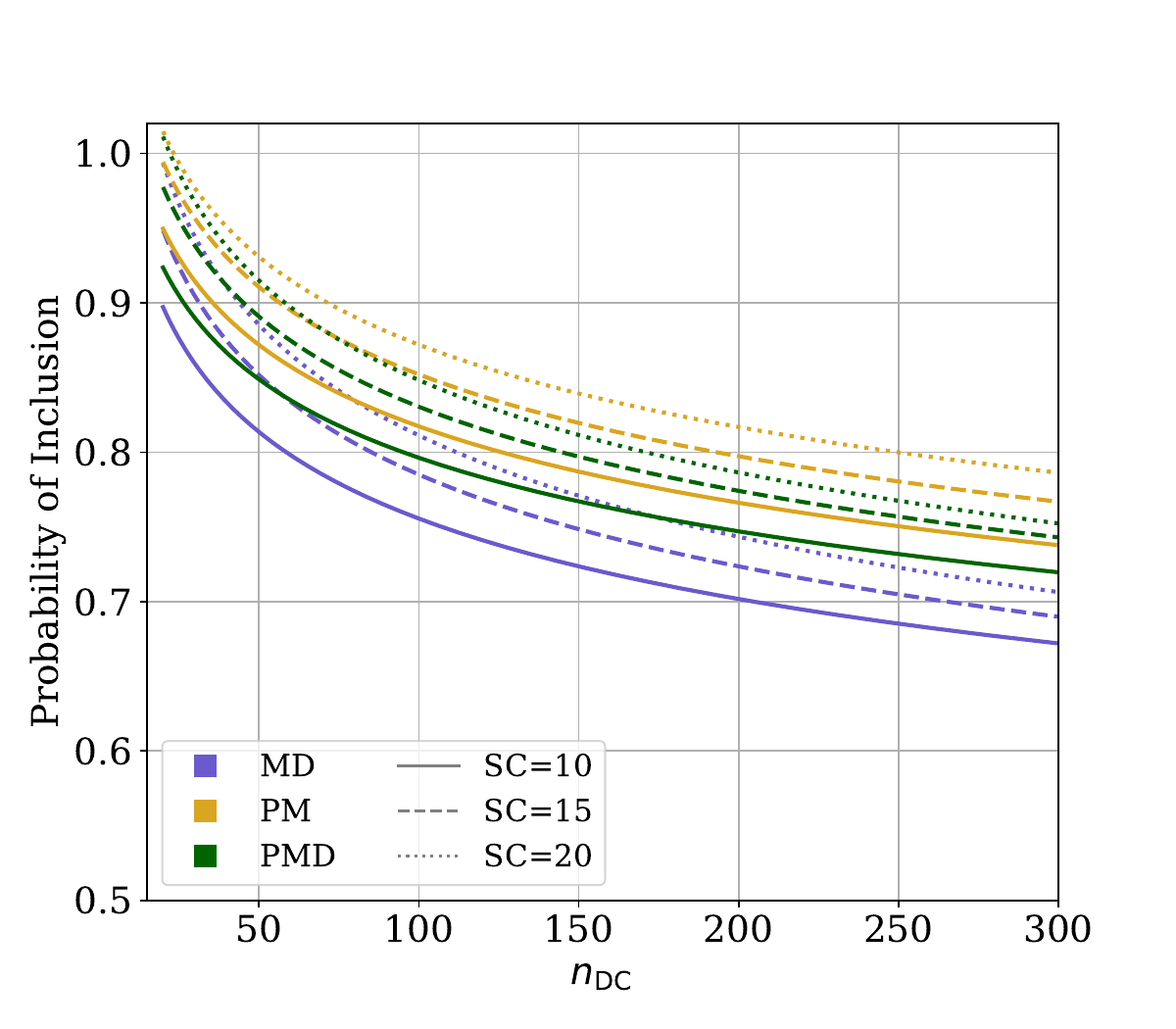}
    \caption{Best-fit probability of inclusion with $n_{\rm DC}$ for various models. The PM model could reach over 80\% \textit{probability of inclusion} when $n_{\rm DC}\lesssim 250$}. 
    \label{fig:inclusion}
\end{figure}

\begin{table}[htbp]
\newcolumntype{C}{>{\centering\arraybackslash}X}
\caption{Best fitting parameters $(a,b)$ for various $n_{\rm SC}$ of three models. }
\label{table:ab}
\begin{tabularx}{\textwidth}{CCCC}
\toprule
$n_{\rm SC}$ & MD                                         & PM                                       & PMD                                      \\ \midrule
10 & 1.2629,  0.1234   & 1.2686,  0.1060 & 1.2723,  0.1060 \\
15 & 1.3745,  0.1332 & 1.3724,  0.1149  & 1.3552,  0.1127 \\
20 & 1.5154,  0.1445 & 1.440,  0.1184 & 1.4362,  0.1165 \\ \bottomrule
\end{tabularx}
\end{table}
For better clarity, we quantify the fraction of data package in which the true KN achieved $\rm rank\leq n_{\rm SC}$ for given $n_{\rm DC}$ as \textit{probability of inclusion}. This metric reflects the classifier's effectiveness in incorporating the true KN among the selected candidates. As expected, the classifier's performance exhibited a declining trend with an increasing amount of $n_{\rm DC}$. Notably, we find a good fitting by a power-law function $P(n_{\rm SC},n_{\rm DC})=a\cdot n_{\rm DC}^{-b}$ with fitting parameters $(a,b)$. The $P(n_{\rm SC},n_{\rm DC})$ curves are depicted  in Figure~\ref{fig:inclusion}. The results for the MD, PM, and PMD models are illustrated in violet, yellow, and green, respectively, and solid, dashed, and dotted lines corresponding to $n_{\rm SC}={ 10,15,20 }$. The optimal fitted parameters are outlined in Table~\ref{table:ab}. Leveraging this fitted curve, we can now estimate the likelihood of capturing true KN in future ToO surveys, thus facilitating our choice of $n_{\rm SC}$. For example, given $\sim 250$ candidates detected and 80\% \textit{probability of inclusion} required, we should characterize $n_{\rm SC}\gtrsim 20$.

\section{Conclusion}\label{sec:conclusion}

In the era of multi-messenger astronomy, especially conducted by LVK and the future $3^{\rm rd}$ generation Cosmic Explorer (CE;~\citet{2019BAAS...51g..35R}) and Einstein Telescope (ET;~\citet{2020JCAP...03..050M}), it will require strongly qualified detectors to be mutually compatible, which necessitates the coordination of numerous facilities to effectively leverage the scientific prospects offered by the upcoming dataset. In light of this, the development of machine learning techniques, integrating physically informed features, becomes crucial to streamline data and minimize the burden of human screening. The negative outcome during LIGO/Virgo O3 run indicates that the AT2017gfo-like KN is abnormal, presenting substantial difficulty in identifying KN systematically to date. And in the $\sim$ 6 months since LIGO/Virgo O4 has been running, many collaborations triggered their ToO observations, e.g. GROWTH\footnote{\url{http://growth.caltech.edu/}} and MASTER GLOBAL Robotic Net\footnote{\url{http://observ.pereplet.ru/}}. To enhance the probability of identifying EM counterparts of GW detections, several brokers have been developed, i.e. KN classifier embedded in \textsc{Fink} broker~\citep{biswas_enabling_2022}, EI-CID which was based on \texttt{RAPID} framework~\citep{chatterjee_-cid_2021}, ZTFReST~\citep{2021ApJ...918...63A}, AleRCE~\citep{2021AJ....161..141S}, Lasair~\citep{2019eeu..confE..51S}, SCiMMA\footnote{\url{https://scimma.org/}} and a fully automated pipeline for KN discovery~\citep{2023arXiv230709213S}. 


In this work, we have presented a KN binary classifier with a modified \texttt{RAPID} framework. Our approach was inspired by the enhancements detailed in \citet{chatterjee_-cid_2021}, where the fine-tuned neural network exhibited promising performance on simulated ZTF data. We start with the simulation of transients where we have conducted mock WFST survey to 60 simulated GW skymaps with O4 sensitivity. The KN are located following the probability distribution of GW skymaps whereas contaminants are located uniformly across the sky. Noted that~\citet{2021ApJ...918...63A} found cosmological afterglow the dominant contaminants at high Galactic latitude within $\sim$ 1 yr observations. Furthermore, other contaminant class, e.g. CVs, afterglows and shock breakouts, poses potential challenges because only a fraction of them can be filtered by cross-matching with catalog, implying it would be dominant sources sometime~\citep{biswas_enabling_2022}. Considering that the classifier is designed to operate on processed data ideally filtered through a real-bogus test and the exclusion of variable stars, we have simulated the majority of supernovae, which represent the predominant contaminants following the removal of these sources. Upon these foundations, we have applied three combinations of contextual information, MD, PM and PMD, revealing comparable performances between the PM and PMD models, while the MD model proves to yield less promising results as shown by the confusion matrix and ROC curve in Figure~\ref{fig:cf} . The PMD model shows comparable accuracy as evident in Figure~\ref{fig:acc} in which accuracy for \textit{Other} reaches $\gtrsim 98\%$ and \textit{Kilonova} reaches $\gtrsim 92\%$ after 3 days since trigger. Beyond that, we also validated models by predicting KN score for Swope observations on AT2017gfo and DECam observations on AT2019npv, where the lightcurves and predicted KN scores are shown in Figure~\ref{fig:real_data}. 

Furthermore, we simulated true KN accompanied by a quantity of contaminants for each mock survey to examine the performance on ToO survey data. By sorting KN scores in descending order, the distribution of rankings of true KN are plotted in Figure~\ref{fig:ranking}, indicating that $\sim 80\%$ probability KN ranking are $\lesssim 20$ when $n_{\rm DC}$ less than 290. In addition, we found a robust fitting to \textit{probability of inclusion} with $n_{\rm SC},n_{\rm DC}$ as shown in Figure~\ref{fig:inclusion} which will instruct the choice of $n_{\rm SC}$ in the future. 

Through the analysis of GW observable, we found a great discrepancy between line-of-sight probability, mean and standard deviation of luminosity distance according to GW. And we did not include offset in cross-matching and $A_{90}$ of GW skymap as in~\citet{chatterjee_-cid_2021} due to the uncertainty in identifying host galaxy and numerical relativity simulations. Alternatively, we explored various KN models and neutron stars, leveraging a comprehensive sampling approach across a BNS sample, in conjunction with the Bulla and MOSFiT models.  We anticipate retraining our models with the inclusion of data from the forthcoming WFST survey, specifically incorporating observations of real supernovae, and expanding the scope of the models to accommodate phenomena like GRB afterglows and stellar flares. Furthermore, we envision fostering collaboration among diverse observational bands, i.e. preliminary X-ray emission detectors like Einstein Probe (EP;~\citet{2015arXiv150607735Y}) or Chandra X-ray Observatory(CXO;~\citet{2000SPIE.4012....2W}) , to glean additional SED features beyond the optical band. 

\vspace{6pt} 




\authorcontributions{Liang R. D. is responsible for transient simulation and collection, training of classifier and investigation for performance; Liu Z. Y. optimizes the ToO survey plan and simulation; code for calculating limiting magnitudes for WFST is provided by Lei L. 
This work is proposed and instructed by Zhao W.}

\funding{This work is supported by the Strategic Priority Research Program of the Chinese Academy of Science (Grant No. XDB0550300), the National Key R\&D Program of China (Grant No. 2021YFC2203102 and 2022YFC2204602), the National Natural Science Foundation of China (Grant No. 12325301 and 12273035), the Fundamental Research Funds for the Central Universities (Grant No. WK2030000036 and WK3440000004), the Science Research Grants from the China Manned Space Project (Grant No.CMS-CSST-2021-B01), the 111 Project for "Observational and Theoretical Research on Dark Matter and Dark Energy" (Grant No. B23042), and Cyrus Chun Ying Tang Foundations.}



\acknowledgments{We deeply thank all the helpful discussion and support by WFST team. We also appreciate constructive suggestions by Lin Z. Y. and Niu R. We are especially grateful to Mourani G. -M. for being involved in the discussion and writing. }

\conflictsofinterest{The authors declare no conflict of interest. } 


\begin{adjustwidth}{-\extralength}{0cm}

\reftitle{References}

\PublishersNote{}
\end{adjustwidth}
\end{document}